\documentclass{nature}

\usepackage{amsmath}
\usepackage[pdftex]{graphicx}
\usepackage{dcolumn}  
\usepackage{bm}       
\usepackage{dsfont}
\usepackage{color}
\usepackage{braket}
\usepackage{amssymb}
\usepackage{booktabs}

\newcounter{defcounter} 
\usepackage{amsthm}

\title{Diverse Transient Chiral Dynamics in Evolutionary distinct Photosynthetic Reaction Centers} 

\author{Yonglei Yang$^{1,*}$, Zihui Liu$^{1,*}$, Fulu Zheng$^{1,2,*}$, Panpan Zhang$^{1}$, Hongxing He$^{1}$, Ajay Jha$^{3,4}$, Hong-Guang Duan$^{1}$ } 

\begin{document} 

\maketitle 

\begin{affiliations} 
\item Department of Physics, School of Physical Science and Technology, Ningbo University, Ningbo, 315211, P.R. China. 
\item Bremen Center for Computational Materials Science, University of Bremen, Am Fallturm 1,28359 Bremen, Germany.
\item Rosalind Franklin Institute, Harwell, Oxfordshire OX11 0QX, United Kingdom. 
\item Department of Pharmacology, University of Oxford, Oxford, OX1 3QT United Kingdom. \\ 
\centerline{\underline{\date{\bf \today}}} 
\end{affiliations} 

\begin{abstract} 

The evolution of photosynthetic reaction centers (RCs) from anoxygenic bacteria to higher order oxygenic cynobacteria and plants highlights a remarkable journey of structural and functional diversification as an adaptation to environmental conditions. The role of chirality in these centers is important, influencing the arrangement and function of key molecules involved in photosynthesis. Investigating the role of chirality may provide a deeper understanding of photosynthesis and the evolutionary history of life on Earth. In this study, we explore chirality-related energy transfer in two evolutionarily distinct reaction centers: one from the anoxygenic purple sulfur bacterium {\em Thermochromatium tepidum} (BRC) and the other from the oxygenic cyanobacterium {\em Thermosynechococcus vulcanus} (PSII RC), utilizing two-dimensional electronic spectroscopy (2DES). By employing circularly polarized laser pulses, we can extract transient chiral dynamics within these RCs, offering a detailed view of their chiral contribution to energy transfer processes. We also compute traditional 2DES and compare these results with spectra related to circular dichroism (CD). Our findings indicate that two-dimensional circular dichroism (2DCD) spectroscopy effectively reveals chiral dynamics, emphasizing the structural symmetries of pigments and their interactions with associated proteins. Despite having similar pigment-protein architectures, the BRC and PSII RC exhibit significantly different chiral dynamics on an ultrafast timescale. In the BRC, the complex contributions of pigments such as $BCh_{M}$, $BPh_{L}$, $BCh_{L}$ and $P_{M}$ to key excitonic states lead to more pronounced chiral features and dynamic behavior. In contrast, the PSII RC, although significantly influenced by $Chl_{D1}$ and $Chl_{D2}$, shows less complex chiral effects and more subdued chiral dynamics. Notably, the PSII RC demonstrates a faster decay of coherence to localized excitonic populations compared to the BRC, which may represent an adaptive mechanism to minimize oxidative stress in oxygenic photosystems. By examining and comparing the chiral excitonic interactions and dynamics of BRC and PSII RC, this study offers valuable insights into the mechanisms of photosynthetic complexes. These findings could contribute to understanding how the functional optimization of photosynthetic proteins in ultrafast timescales is linked to biological evolution.

\end{abstract} 


Chirality is a fundamental principle in chemistry, essential for the function of biomolecules and central to the evolution of life on Earth. It refers to the symmetry properties of the system, characterized by the absence of mirror planes or inversion symmetry. Consequently, a chiral system and its mirror image cannot be superimposed through simple rotations or translations. In living organisms, chirality affects biochemical interactions, enzyme specificity, and the structural stability of macromolecules, including proteins and nucleic acids. The preference for specific chiral forms in biological systems suggests a deep evolutionary advantage that has been refined over billions of years\cite{Chirality1, Chirality2, Chirality3}. 

Photosynthesis, the process of converting light energy into chemical energy, exemplifies the importance of chirality in biological functions. Over billions of years, the evolution of photosynthesis has resulted in the diversification and specialization of photosynthetic organisms, from simple anoxygenic bacteria to more complex oxygenic cyanobacteria and plants \cite{Blankenship book, Photosynth_res_evolution}. Central to this evolutionary process are photosynthetic reaction centers (RCs), which are the primary sites where sunlight is transformed into chemical energy. In photosynthetic RCs complexes, structural fluctuations in the protein framework can be studied using differential chiral optical responses following light absorption, as the extent of exciton delocalization dynamically changes. Examining the role of chirality by studying these transient chiral optical responses in evolutionary distinct RCs (oxygenic PSII RC\cite{Nature PSII structure 1, Nature PSII structure 2, Nature PSII structure 3} and anoxygenic bacterial RC\cite{Blankenship book, BRC structure 1, BRC structure 3}) will offer a deeper understanding towards the structural and functional optimization of photosynthetic proteins associated with the biological evolution.

Transient optical responses associated with ultrafast energy and charge transfer have been widely studied using various time-resolved spectroscopic techniques. Rienk van Grondelle and colleagues investigated ultrafast dynamics within the PSII reaction centers of oxygenic photosystems\cite{Grondelle 1}. They used transient absorption spectroscopy to explore energy transfer and primary charge-transfer pathways\cite{Grondelle 2}. By extending the temporal probing range, they observed charge separation dynamics occurring up to 3 nanoseconds\cite{Grondelle 3}. Complementing this approach, Novoderezhkin et al. employed excitonic modeling to compute absorption and other linear spectra, developing a model with refined parameters by fitting to experimental spectra at various temperatures\cite{Vladimer 1, Vladimer 2, Vladimer 3}. Two-dimensional electronic spectroscopy (2DES) has been very useful to spectrally resolve ultrafast transient optical responses in photosynthetic complexes\cite{Fuller, Cao, Duan1, Duan2, Zig_NatChem1, Zig_NatChem2, Jha_SciAdv, Zhou_Phot}. Ogilvie and colleagues utilized 2DES to study excitonic dynamics and associated coherences in PSII reaction centers\cite{Jennifer JPCL}. With its superior time resolution and enhanced signal-to-noise ratio, 2DES enabled the resolution of both electronic and vibrational coherences\cite{Jennifer NC 2014}. Additionally, using a multi-color configuration, they investigated energy and charge transfer dynamics through visible and infrared spectra, identifying low-energy excited states characteristic of charge-transfer states\cite{Jennifer Nature Comm 1, Jennifer Science Adv}. Beyond oxygenic systems, 2DES has also been employed to study reaction centers in anoxygenic bacteria\cite{Grondelle NC, Zigmantas SA}. Multi-color 2DES has further facilitated the investigation of primary charge-transfer pathways\cite{Jennifer PNAS}. Studies using 2DES have reported the existence of long-lived vibronic coherence\cite{Jennifer Science advances}, underscoring its efficacy in capturing transient optical responses in photoexcited RCs.

To explore chiral optical responses, the use of circularly polarized light has been proposed for studying energy transfer and transient chiral dynamics within photosynthetic protein complexes. Recent studies have demonstrated that 2DES can be adapted to detect chiral signals, which reveal how excitonic states in light-harvesting complexes delocalize and contract dynamically following photoexcitation\cite{Engel NC 2014}. This capability makes 2DES a powerful technique for investigating ultrafast molecular dynamics in chiral environments, providing insights into energy transfer mechanisms that are not accessible through conventional non-chiral methods. Supporting theoretical works by Cho {\em et al.} and Duan {\em et al.} have further validated these findings\cite{Cho JCP 2007, Duan PRE 2023}. Recent studies on FMO complexes have demonstrated that using two-dimensional circular dichroism (2DCD) spectroscopy is promising for detecting chiral optical responses in photosynthetic complexes.\cite{Liu_JPCL}. Thus, 2DCD spectroscopy emerges as a valuable tool for probing transient chiral responses in photosynthetic reaction centers.

In this study, we investigate the transient chiral dynamics and associated quantum coherences in photosynthetic RCs, from simpler anoxygenic bacteria {\em Thermochromatium tepidum} to more complex oxygenic cyanobacteria {\em Thermosynechococcus vulcanus}. {\em T. tepidum} is part of the Proteobacteria, which performs anoxygenic photosynthesis. It uses sulfur compounds as electron donors and has a simpler photosynthetic apparatus (based on bacteriochlorophyll). {\em T. vulcanus} belongs to the Cyanobacteria, which are considered one of the earliest groups to perform oxygenic photosynthesis, contributing to the Great Oxygenation Event in Earth's history\cite{Lyons_Nature_2014}. Both {\em T. tepidum} and {\em T. vulcanus} are thermophiles, adapted to thrive in high-temperature environments, suggesting the possibility of convergent evolution as a result of similar environmental pressures. The evolution of photosynthesis indicates that these organisms likely diverged early in the evolutionary timeline, with cyanobacteria developing a more complex and efficient oxygenic photosynthetic system. Utilizing circularly polarized light in a multi-pulse configuration, we detect chirality-related energy transfer and coherent dynamics within these RCs. Our models and parameters are refined through simultaneous fitting to absorption and CD spectra. Our findings reveal that, due to the symmetrical arrangement of cofactors, the lowest exciton energy states exhibit the strongest chiral signals in the 2DCD spectra of the PSII RC complex from {\em T. vulcanus}. We identify chirality-related cross peaks in the 2DCD spectra during downhill energy transfer. Furthermore, we examine the chiral dynamics in the RC complex of {\em T. tepidum} (BRC). By comparing 2DCD data with 2DES results, we demonstrate that PSII RC and BRC display fundamentally different chiral dynamics during energy transfer, despite having similar pigment-protein structural arrangements. These comparative studies of PSII RC and BRC offer new perspectives on the evolution and diversification of ultrafast photophysical processes in RCs. 


\section*{Results}  

\subsection{Steady state absorption and CD spectra of the PSII RC and BRC complexes} 

\begin{figure}[h!]
\begin{center}
\includegraphics[width=16.0cm]{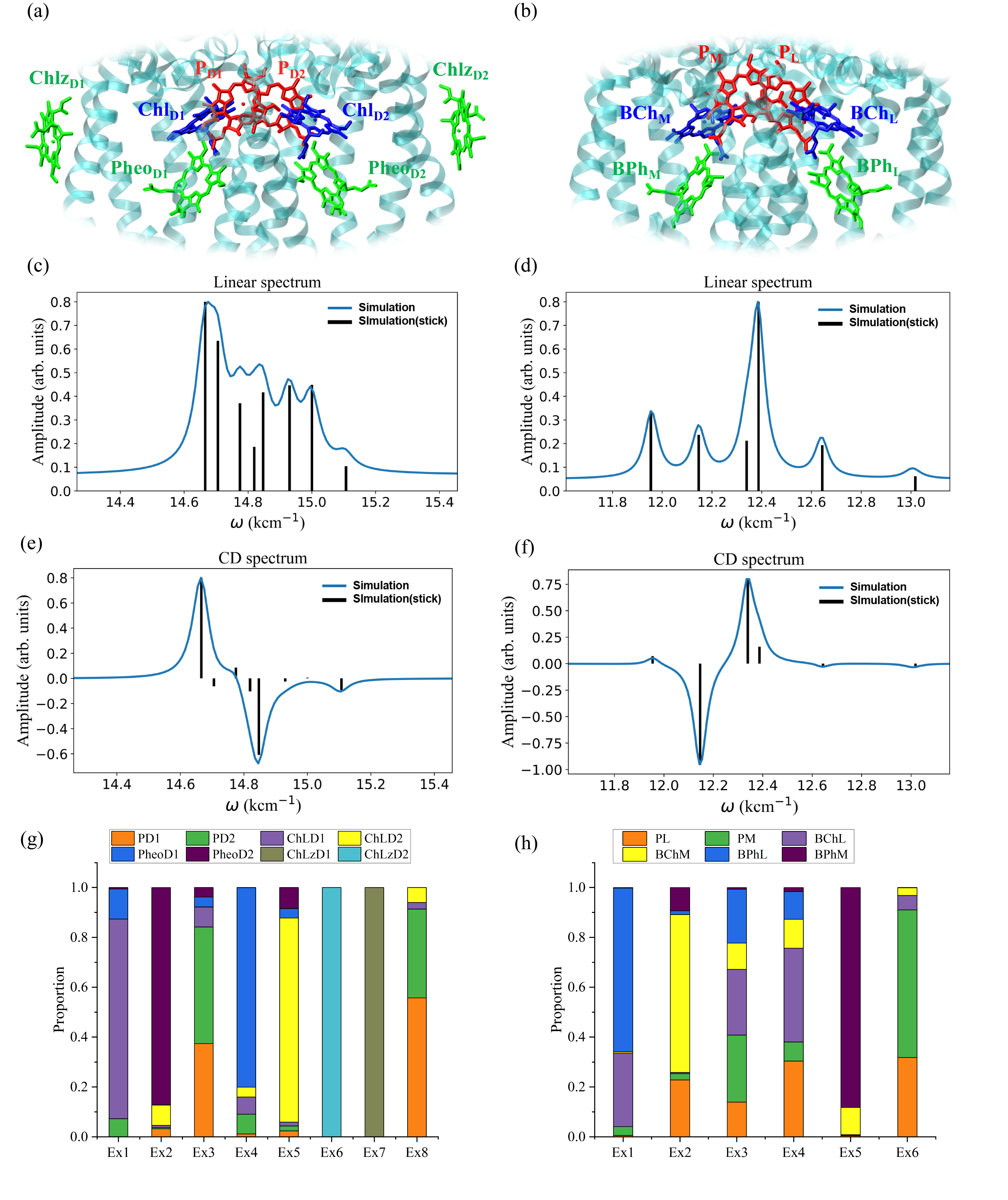} 
\caption{\label{fig:Fig1} (a) arrangement of pigments in PSII RC complex from cyanobacteria {\em Thermosynechococcus vulcanus}. (b) pigments arrangement in BRC from {\em Thermochromatium tepidum}. The linear absorption and CD spectra of PSII RC are presented in (c) and (e), respectively. The absorption and CD spectra of BRC are shown in (d) and (f), respectively. The contributions of pigments to different excitons are plotted in (g) and (h) for PSII RC and BRC, respectively. } 
\end{center}
\end{figure}

The protein structures and pigment arrangements of the PSII RC from cyanobacteria {\em T. vulcanus} and BRC from simpler anoxygenic bacteria {\em T. tepidum} are depicted inFig.\ \ref{fig:Fig1}(a) and (b), respectively. Pigments and their symmetrical partners are labeled using identical colors. In Fig.\ \ref{fig:Fig1}(a), the special pair (${P_{D1}}$ and $P_{D2}$) is highlighted in red, while the other pigments are represented in green and blue. The absorption and CD spectra for the PSII RC are computed and illustrated in Fig.\ \ref{fig:Fig1}(c) and (e), respectively. The excitonic energy levels of the PSII RC are presented as stick spectra in these figures. The broadened absorption and CD spectra are shown as blue solid lines. Excitonic models have been constructed and parameters refined to match these spectra, with detailed model descriptions and parameter values provided in the Methods section. Fig.\ \ref{fig:Fig1}(e) illustrates the CD spectrum, characterized by pronounced positive and negative peaks at the first and fifth sticks ($Ex_{1}$ and $Ex_{5}$), closely linked to the chiral configuration of the pigments. To further elucidate the relationship between excitons and pigments, we calculated and plotted the contributions of each excitonic stick to the PSII RC in Fig.\ \ref{fig:Fig1}(g). This analysis allows us to disentangle the primary contributors to the chiral signals observed in the CD spectrum. Notably, Fig.\ \ref{fig:Fig1}(g) reveals that the $Ex_{1}$ and $Ex_{5}$ are predominantly centered to $Chl_{D1}$ and $Chl_{D2}$, respectively, both of which are symmetrically positioned within the PSII RC protein structure. Moreover, $Chl_{D1}$ exhibits the lowest site energy within the protein complex, making it the initial trapping site for population transfer following photoexcitation. Tracking the temporal evolution of the first and fifth excitonic peaks in the CD spectrum can thus provide insights into the time-resolved chiral dynamics associated with energy and charge transfer in the PSII RC.

Using a similar approach, we present the pigment configuration of the BRC in Fig.\ \ref{fig:Fig1}(b). The corresponding calculated absorption and CD spectra are shown in Fig.\ \ref{fig:Fig1}(d) and (f). The stick spectra are depicted as black solid bars, while the broadened calculated absorption spectra are shown as blue solid lines. Contributions from each stick are illustrated in Fig.\ \ref{fig:Fig1}(h), enabling us to identify the major contributors to the CD spectrum. For example, the second and third sticks ($Ex_{2}$ and $Ex_{3}$), which exhibit the highest amplitudes in Fig.\ \ref{fig:Fig1}(h), are primarily associated with $BCh_{M}$, with the third stick ($Ex_{3}$) involving a more complex combination of pigments, including $P_{M}$, $BCh_{L}$, and $BPh_{L}$. Analyzing the time-resolved dynamics of these sticks in the CD spectra and their roles in downhill energy transfer can be further investigated using 2DCD spectroscopy, which will be discussed in the next section.

\subsection{Chiral dynamics of PSII RC complex from cyanobacteria {\em Thermosynechococcus vulcanus}} 

\begin{figure}[h!]
\begin{center}
\includegraphics[width=12.0cm]{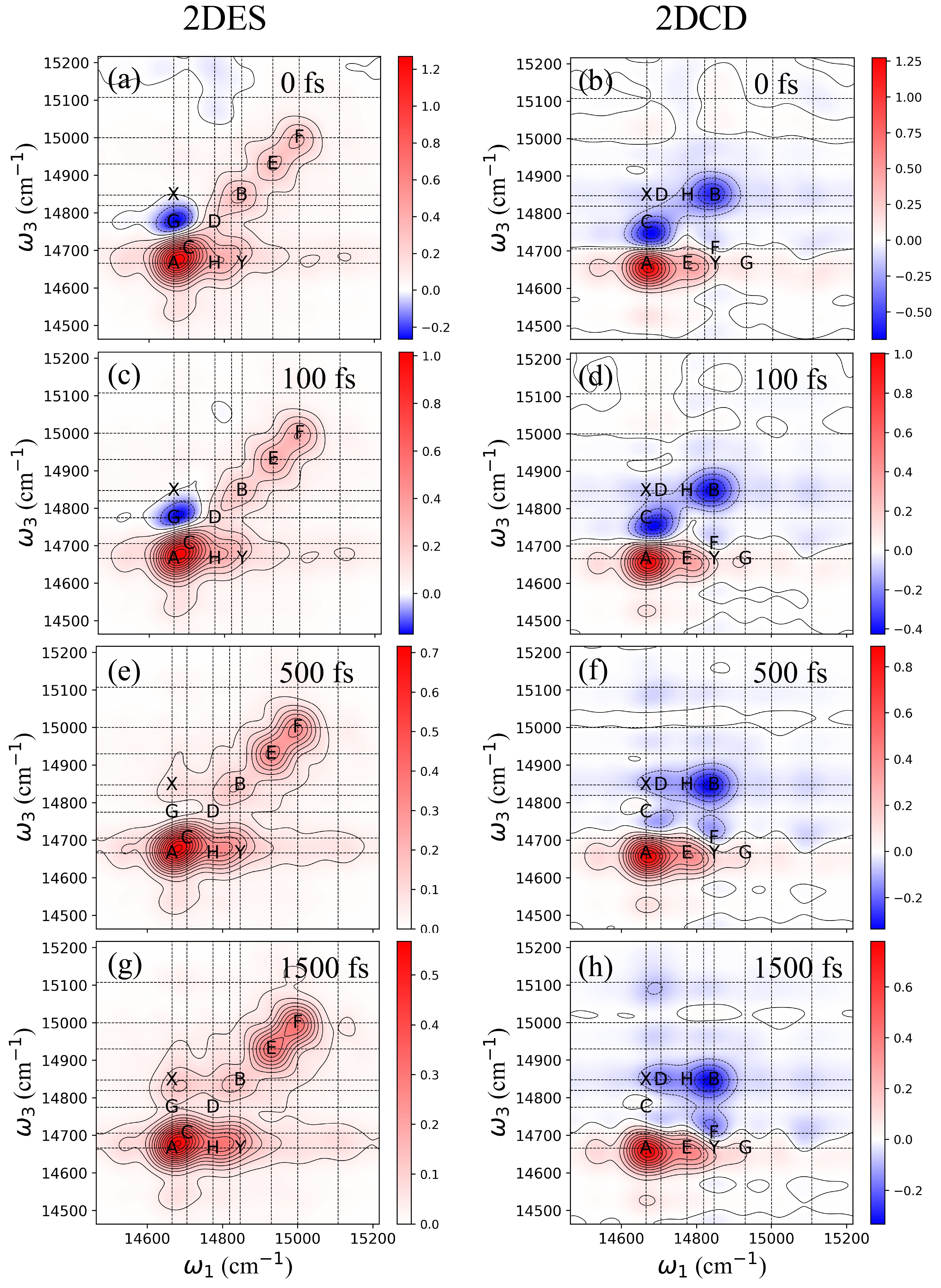}
\caption{\label{fig:Fig2} The 2D electronic spectra (2DES) of the PSII reaction center from {\em T. vulcanus} at selected waiting times of 0, 100, 500, and 1500 fs are displayed in panels (a), (c), (e), and (g), respectively. Corresponding 2D circular dichroism (2DCD) spectra for the same waiting times are shown in panels (b) through (h). In these spectra, positive amplitudes indicate contributions from ground-state bleaching and stimulated emission, while negative magnitudes correspond to optical signals arising from excited-state absorption. These features provide insight into the underlying photophysical processes and their evolution over time.} 
\end{center}
\end{figure}

In this section, we present the calculated 2DES and 2DCD spectra of PSII RC at selected waiting times of 0, 100, 500, and 1500 femtoseconds (fs). Fig.\ \ref{fig:Fig2}(a), (c), (e), and (g) show the 2DES results, with solid crosses marking the positions of eigenstates in the PSII RC to provide better resolution of the dynamics. Population dynamics calculations were carried out using the Redfield quantum master equation\cite{Redfield1, Redfield2}, while the 2DES and 2DCD spectra were computed using response function theory\cite{Mukamel's book}. The specific calculation details are described in the previous section and in the Supplementary Information (SI).

Fig.\ \ref{fig:Fig2}(a) illustrates the 2DES of the PSII RC, displaying diagonal peaks labeled A through F. Cross peaks with both positive and negative magnitudes are also observed and marked as G, H, X, and Y, respectively. Notably, the blue peak G exhibits a strong negative value, indicating the presence of excited-state absorption (ESA). This ESA peak decays within 400 fs. As time progresses, downhill energy transfer becomes evident, highlighted by the increasing intensity of cross peaks (H, Y) and other peaks located in the bottom-right region. In contrast, the 2DCD spectra, shown in Fig.\ \ref{fig:Fig2}(b), (d), (f), and (h) for the same waiting times, reveal different characteristics. Solid crosses again indicate excitonic energy levels, with main and cross peaks labeled A through H. Peaks A and B exhibit both positive and negative magnitudes, corresponding to the main peaks in the CD spectrum shown in  Fig.\ \ref{fig:Fig1}(e). Interestingly, other excitonic states do not produce strong signals in the CD and 2DCD spectra. Analyzing the contributions from the stick spectrum in  Fig.\ \ref{fig:Fig1}(g) allows us to correlate the 2DCD spectral signals with the chiral arrangement of pigments within the PSII RC. Notably, peaks A and B, which are the most prominent in the 2DCD spectra, correspond to the first and fifth excitonic states ($Ex_{1}$ and $Ex_{5}$) and are primarily influenced by the pigments $Chl_{D1}$ and $Chl_{D2}$. These pigments exhibit a symmetric arrangement along the central axis of the PSII RC, underscoring the strong relationship between these main peaks and the cofactors $Chl_{D1}$ and $Chl_{D2}$. Furthermore, cross peaks are observed between A and C in the 2DCD spectra, with their magnitudes (X and Y) changing over time (as detailed in the next section). The upper-left cross peak X initially shows weak intensity, which increases over time. Previous studies indicate that population transfer to the lowest excitonic state, $Chl_{D1}$, occurs over time, leading to an imbalance between the symmetric pigments $Chl_{D2}$ and $Chl_{D1}$, thereby causing a significant increase in the magnitude of cross peak X. Although the bottom-right peak Y could also provide this spectroscopic information, it is affected by downhill energy transfer from other pigments.

Next, we analyze the transient chiral dynamics between the special pair $P_{D1}$ and $P_{D2}$ in the PSII RC. The close proximity of these pigments (3 $\AA$) results in strong excitonic coupling, approximately 150 cm$^{-1}$, leading to significant mixing in the excitonic basis. As indicated in Fig.\ \ref{fig:Fig1}(g), the third and eighth excitonic states ($Ex_{3}$ and $Ex_{8}$) are predominantly contributed by this special pair. Despite the strong excitonic interaction, cross peaks and associated electronic coherence between these states are absent in the 2DCD spectra and appear weak in the 2DES, possibly due to the large excitonic energy gap. This suggests that main and cross peaks in the 2DCD spectra are not directly related to excitonic coupling between pigments at the site basis but are more closely associated with the chiral arrangement of pigments.

We further investigate the population dynamics and transient chiral signals between the two Pheophytins, $Pheo_{D1}$ and $Pheo_{D2}$. Analyzing their contributions in the excitonic stick spectrum reveals that the second and fourth excitonic states ($Ex_{2}$ and $Ex_{4}$) are involved. However, these states do not generate significant main or cross peaks in the 2DCD spectra. Additionally, we explore the chiral dynamics between the side pigments, $Chlz_{D1}$ and $Chlz_{D2}$. The substantial distance between these pigments makes strong excitonic coupling unlikely, resulting in monomer-like absorption behavior, as evidenced by their contributions to the excitonic states in the stick spectrum. The sixth and seventh excitonic energy levels ($Ex_{6}$ and $Ex_{7}$) are mainly associated with these side pigments, yet no prominent main or cross peaks are observed between these levels in the 2DCD spectra. These findings demonstrate that 2DCD spectroscopy is not significantly influenced by strong excitonic coupling between pigments. Moreover, electronic transitions of pigments do not play a dominant role in shaping the 2DCD spectra. Instead, the chiral arrangement of pigments, in conjunction with other cofactors, primarily determines the spectroscopic signals observed in the 2DCD spectra.

\subsection{Comparative kinetic analysis of 2DES and 2DCD data}  

\begin{figure}[h!]
\begin{center}
\includegraphics[width=16.0cm]{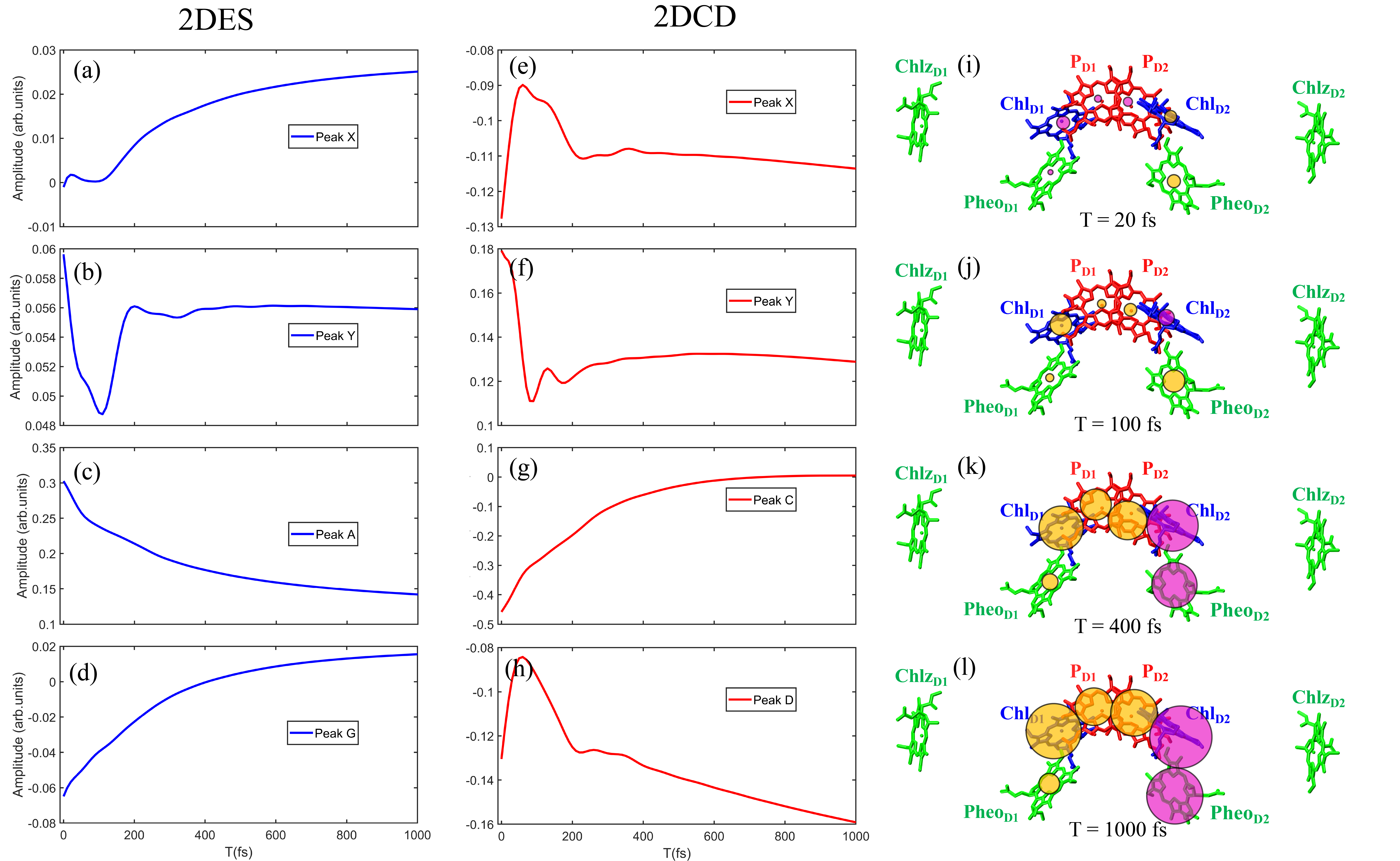} 
\caption{\label{fig:Fig3} Kinetic traces of the selected diagonal excitonic peaks, along with the cross peaks in 2DES and 2DCD of the PSII reaction center, are presented over waiting times ranging from 0 to 1000 fs. The names of the peaks, as defined in Fig.\ \ref{fig:Fig2}, are labeled in each plot. Chiral dynamics associated with population transfer are shown in panels (i) to (l) for selected waiting times of T = 20, 100, 400, and 1000 fs. The chirality-related variations are compared to the magnitudes of 2DCD at T = 0 fs, with positive and negative magnitudes represented by magenta and yellow circles, respectively.}
\end{center}
\end{figure}

In this section, we present the traces and kinetics of selected diagonal and cross peaks in the 2DES, as illustrated in Fig.\ \ref{fig:Fig3}. The positions of these peaks are marked in Fig.\ \ref{fig:Fig2}. We begin by analyzing the cross peaks 'X' and 'Y', which correspond to the interactions between the first and fifth excitonic states ($Ex_{1}$ and $Ex_{5}$). To investigate these dynamics, we extract the amplitude of these peaks and plot their time-resolved behavior in Fig.\ \ref{fig:Fig3}. Notably, as the ESA decays, the magnitude of the X peak significantly increases over time, reaching its maximum at approximately 1000 fs. In contrast, the Y peak exhibits a rapid decay in magnitude, stabilizing at equilibrium after T $>$ 200 fs. We also depict the time-resolved magnitudes for peaks A and G in Fig.\ \ref{fig:Fig3}, with additional traces for other selected peaks available in the SI. Evidence of electronic coherence is clearly observed in the peaks labeled (d) and (e) presented in the SI. Compared to the 2DES, the kinetics of peaks in the 2DCD spectra exhibit slightly different behavior. The traces for cross peaks X and Y are plotted, revealing distinct kinetics compared to their 2DES counterparts. Specifically, the peak X shows an initial rapid decay in magnitude, followed by a subsequent increase with waiting time, continuing even beyond 1 ps. Additionally, the kinetic trace of Y peak display a rise in magnitude, peaking at around 1 ps. As discussed in the previous section, the cross peaks X and Y are predominantly influenced by the pigments $Chl_{D1}$ and $Chl_{D2}$. Based on the long-term calculations of the 2DCD spectra, we observe a decay in chirality-related signals, which is closely associated with the structural arrangement of $Chl_{D1}$ and $Chl_{D2}$ relative to other pigments within the PSII RC. Interestingly, the peak magnitudes do not rapidly decay within 1 ps and may persist for an extended duration, even though the population predominantly resides in the lowest energy site of $Chl_{D1}$. Moreover, oscillatory dynamics are evident in the traces of the X and Y peaks. It is noteworthy that the kinetics of peak Y appears to be less affected by energy transfer processes. However, the traces of peaks in Fig.\ \ref{fig:Fig3}(g) and (h) show no distinct evidence of electronic coherence compared to the 2DES, with further details provided in the SI. 

The transient chiral dynamics of excitons can be revealed by an asymmetric population distribution between two pigments, such as the radical pair ($P_{D1}$ and $P_{D2}$). An increase in the population difference between these pigments leads to an enhancement of transient chiral signals in the 2DCD spectrum. Initially, it is assumed that the population is evenly distributed across the pigments following photoexcitation. Consequently, the population difference between symmetrically arranged pigments is minimal, resulting in a relatively weak chiral signal in the 2DCD spectrum at T = 0 fs, as depicted in Fig.\ \ref{fig:Fig3}(i). As time progresses, cross peaks emerge between the first and fifth excitonic states ($Ex_{1}$ and $Ex_{5}$), indicating an increasing asymmetry in population distribution between these excitons over time. This suggests that the chiral signal, driven by population differences, is becoming more pronounced among the delocalized wave packets. To illustrate the chirality-related dynamics from population difference, we further extract the magnitude of related cross peaks in 2DCD spectra and plot the chirality-related population difference from Fig.\ \ref{fig:Fig3}(i) to (l). We use the diameter of circles to present the strengths of chiral signal, which relates to the population difference. Moreover, we use the amplitudes of initial 2DCD spectrum (T = 0 fs) as a benchmark and plot the positive (negative) magnitude as magenta (yellow) circles. The details of this analysis process have been described in the SI. Through this exciton-site transformation analysis, it is evident that the delocalized wave packet predominantly resides in $Chl_{D1}$ and $Pheo_{D1}$.  As time evolves (T $>$ 600 fs), the transient chiral signals stabilize between the first and fifth excitonic states, indicating that this persistent chiral dynamic primarily originates from the interaction between $Chl_{D1}$ and $Chl_{D2}$. This behavior is attributed to the population difference between these two pigments, with $Chl_{D1}$ having the lowest site energy, thereby sustaining the observed chiral signal.

\subsection{Chiral dynamics of BRC complex from {\em Thermochromatium tepidum}} 

\begin{figure}[h!]
\begin{center}
\includegraphics[width=13.0cm]{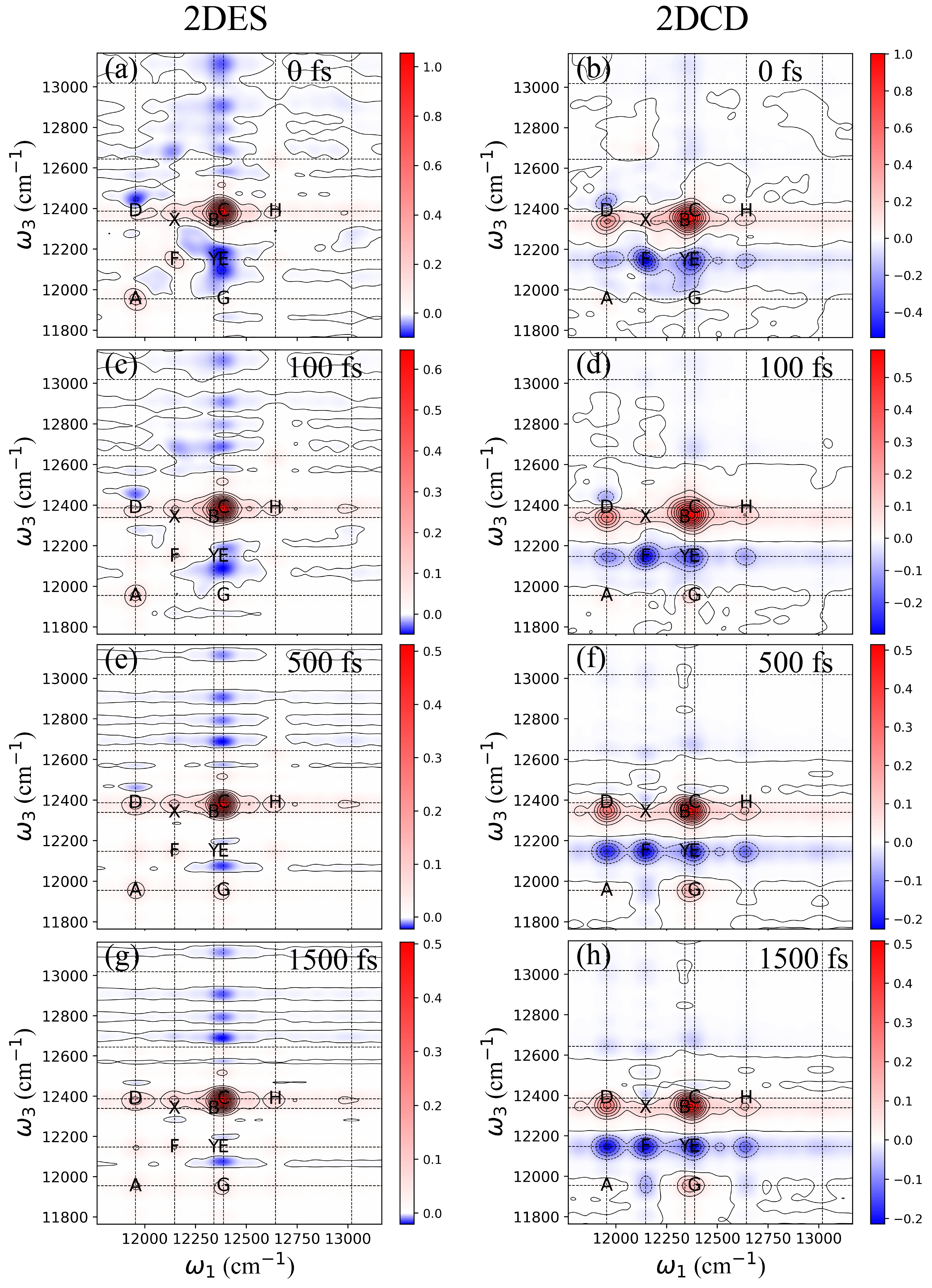}
\caption{\label{fig:Fig4}The 2DES and 2DCD spectra of the bacterial reaction center (BRC) from {\em T. tepidum} are presented for selected waiting times of 0, 100, 500, and 1500 fs. In these spectra, key diagonal excitonic peaks and cross peaks are labeled alphabetically, ranging from A to Y. These peaks represent significant features within the excitonic landscape, providing insight into the dynamics of energy transfer and the interplay between excitonic states during the time evolution.} 
\end{center}
\end{figure}

In this section, we present the calculated 2DES and 2DCD spectra of the BRC, as illustrated in Fig.\ \ref{fig:Fig4}. The 2DES spectra are shown in Fig.\ \ref{fig:Fig4}(a), (c), (e), and (g) for waiting times of 0, 100, 500, and 1500 fs, respectively. The digonal and cross peaks in the BRC spectra are more distinctly separated compared to those in the PSII RC spectra shown in Fig.\ \ref{fig:Fig2}, despite the structural similarities between BRC and PSII RC proteins. Excitonic energy levels are marked with crosses on the 2DES spectra. The diagonal peaks are marked as A, F, B, and C, alongside cross peaks E, G, H and D in the 2DES spectra. Notably, peak E, characterized by a strong negative (blue) signal, is an ESA feature. The temporal evolution of these peaks will be further analyzed in the subsequent section using time-resolved traces. 

For comparative analysis, we have also calculated the 2DCD spectra, displayed in Fig.\ \ref{fig:Fig4}(b), (d), (f), and (h) at the same waiting times (0, 100, 500, and 1500 fs). The 2DCD spectra exhibit well-separated diagonal and cross peaks, also labeled from A to H. Based on CD spectra and the pigment contributions to excitonic states, it is evident that two prominent CD signals predominantly originate from the second and third excitonic states ($Ex_{2}$ and $Ex_{3}$). Accordingly, we have marked the cross peaks between the second and third excitonic states ($Ex_{2}$ and $Ex_{3}$) as X and Y in both 2DES and 2DCD spectra. The 2DCD spectra, especially at extended waiting times (T $>$ 1 ps), provide clearer evidence of ESA peaks than the 2DES spectra. Analysis of pigment contributions shows that bar 2 ($Ex_{2}$) is mainly associated with $BCh_{M}$, whereas bar 3 ($Ex_{3}$) involves contributions from $BPh_{L}$, $BCh_{L}$ and $P_{M}$. This finding suggests that in the 2DCD spectra, peaks F and B are predominantly associated with $BCh_{M}$ and a combination of $BPh_{L}$, $BCh_{L}$ and $P_{M}$, respectively. The cross peaks X and Y reflect the chiral organization of these pigments relative to others within the BRC. Interestingly, peak D in the 2DCD spectra demonstrates a relatively stronger intensity than in the 2DES spectra, implying that the combined electronic and magnetic transition dipole moments bring a stronger influence than pure electronic transitions alone. This observation is rare in nature and likely results from the unique structural arrangement of pigments and proteins. Additionally, the increasing intensity of cross peak G provides evidence of downhill energy transfer within the BRC.

\begin{figure}[h!]
\begin{center}
\includegraphics[width=16.0cm]{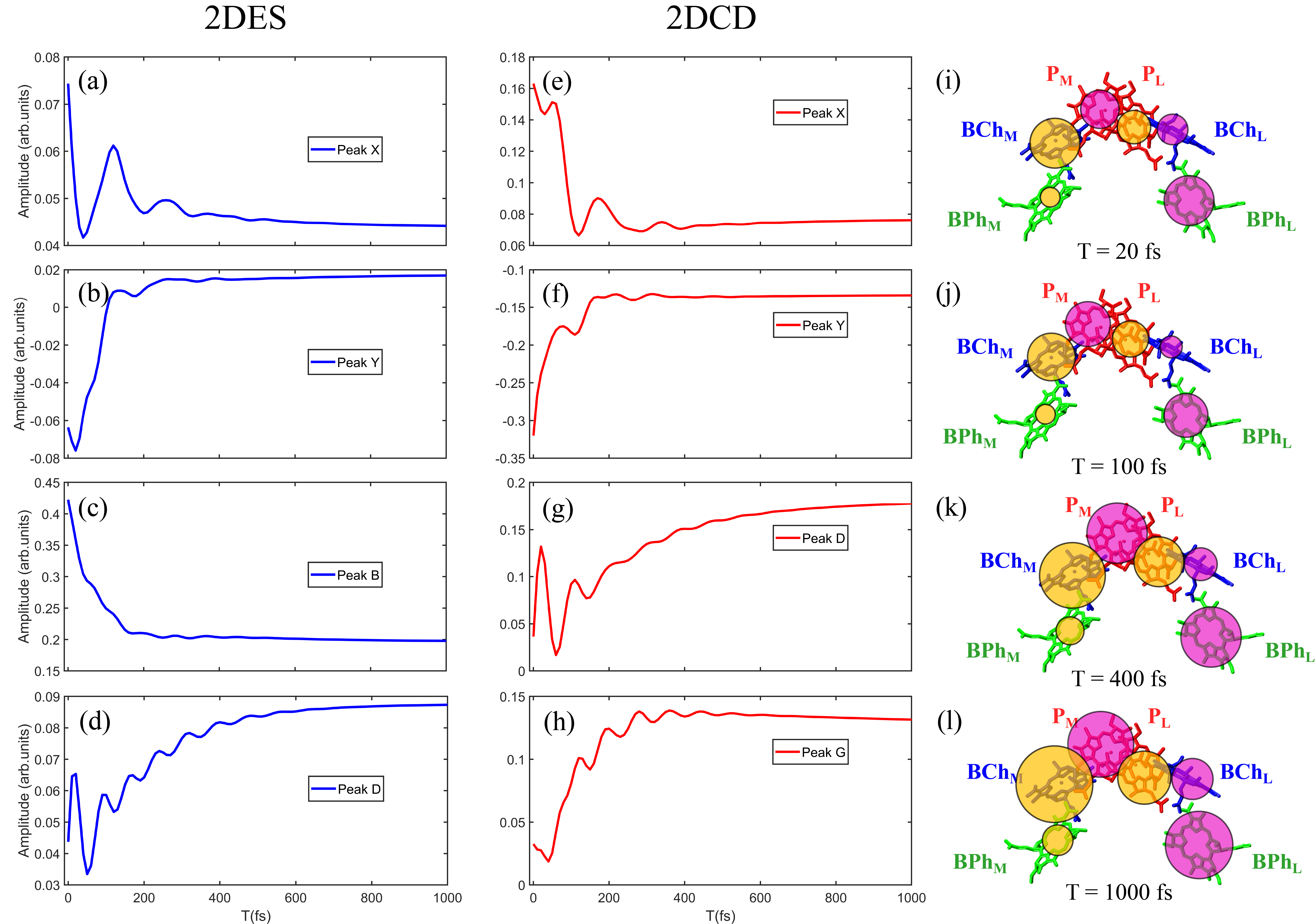}
\caption{\label{fig:Fig5}(a-h) Kinetic traces of selected peaks from the 2DES and 2DCD spectra of the BRC are presented, with the specific peaks highlighted in Fig.\ \ref{fig:Fig4}. The chiral dynamics associated with population differences are depicted in panels (i) to (l) for selected waiting times of 20, 100, 400, and 1000 fs, respectively. These traces provide a detailed view of the temporal evolution of chiral excitonic interactions and population transfer dynamics within the system.} 
\end{center}
\end{figure}

We proceed by examining the population dynamics in both 2DES and 2DCD spectra. For this purpose, we have extracted the magnitudes of selected peaks and plotted their time-resolved dynamics in Fig.\ \ref{fig:Fig5}. The kinetics of cross peaks X and Y are shown as blue solid lines. Peak X exhibits a pronounced oscillatory behavior, decaying within 600 fs, while peak Y shows a marked decrease in intensity, stabilizing around 200 fs. Fig.\ \ref{fig:Fig5}(c) to (d) depict the kinetics of other peaks, highlighting clear oscillatory signatures that underscore strong excitonic coupling among the pigments in BRC. Additional peak dynamics are presented in the SI. The time-resolved amplitudes of the 2DCD spectra are also plotted in Fig.\ \ref{fig:Fig5}, with peaks X and Y shown in (e) and (f), respectively. These traces indicate that aside from oscillatory dynamics, the peak magnitudes reach their maximum at approximately 200 fs, corresponding to the timescale observed in the population dynamics of 2DES peaks X and Y in Fig.\ \ref{fig:Fig5}(a) and (b), respectively. Beyond the coherent dynamics represented by oscillatory traces, it is predicted that the chiral arrangement involving $BCh_{M}$ and ($BPh_{L}$, $BCh_{L}$ and $P_{M}$) can be distinctly identified during the temporal evolution, with these signals persisting for extended periods ($>$ 1 ps). Moreover, we analyzed the kinetics of selected peak D and compared it with the 2DES results, showing its trace in Fig.\ \ref{fig:Fig5}(d). The oscillatory nature of the trace highlights the role of excitonic coupling and resultant electronic coherence. The increasing intensity of this trace suggests a significant rise in chiral interactions involving populations between the first and fourth excitonic states ($Ex_{1}$ and $Ex_{4}$). Calculations indicate that these states are mainly associated with contributions from $P_{M}$, $P_{L}$, and $BCh_{L}$. Finally, Fig.\ \ref{fig:Fig5}(h) presents the kinetics of peak G, which also exhibit oscillatory dynamics. The information of time-evolved chiral population of BRC is depicted in Fig.\ \ref{fig:Fig5}(i) to (l) with selected waiting times of 20, 100, 400 and 1000 fs. It is interesting to note that, in BRC, the chirality-related population difference rapidly increases even at early waiting time (T = 20 fs in Fig.\ \ref{fig:Fig5}(i)). Also, the chirality-related population difference shows significant larger values (depicted by larger diameter of circles), which is due to the strong excitonic couplings between pigments in the BRC. Compare to the results in PSII RC (as shown in Fig.\ \ref{fig:Fig3}(i-l)), we observe dramatic difference of this population-transfer-induced chiral dynamics.

\section*{Discussion} 

The contributions of specific pigments to excitonic states play a crucial role in defining the spectral and dynamic properties of both BRC and PSII RC. In BRC, complex pigment contributions (e.g., $BCh_{M}$, $BPh_{L}$, $BCh_{L}$ and $P_{M}$) to key excitonic states result in pronounced chiral features and dynamic behaviors. On the other hand, in PSII RC, although $Chl_{D1}$ and $Chl_{D2}$ contribute significantly to the chiral signals, their effects are less complex, and the observed chiral dynamics are lowered. The analysis of chiral dynamics reveals that BRC maintains stronger chiral signatures and exhibits extended coherence times of $>$ 1 ps. This prolonged coherence indicates a more stable and robust chiral environment within BRC, which is influenced by the unique structural organization of its pigments. In contrast, PSII RC shows a faster decay of coherence, likely due to the symmetric but less convoluted (in terms of exciton delocalisation) arrangement of its key pigments, like $Chl_{D1}$ and $Chl_{D2}$. This could be a result of evolutionary adaptations to high-intensity light environments, where rapid energy dissipation are crucial to prevent photodamage. The less distinct chiral dynamics and faster loss of coherence in PSII RC might reflect a trade-off between stability and the need to quickly cycle between energy absorption, transfer, and dissipation processes.

The observed differences in chiral dynamics between PSII RC and BRC can also be understood in terms of protection mechanisms that are required because of the highly oxidising chemistry associated with water oxidation in PSII RC. Being oxygenic, {\em T. vulcanus} produces oxygen as a by-product of photosynthesis, which can lead to the generation of reactive oxygen species (ROS) under high light conditions. To protect against oxidative damage, these cyanobacteria have evolved robust photoprotection and repair mechanisms. The less pronounced chiral dynamics and shorter coherence times leading to more more localized excitons in the PSII RC might also be one of these protective adaptations to minimize oxidative stress. When energy is confined to localized states, it is easier for the system to channel this energy into quenching pathways that safely dissipate it. This helps in preventing the transfer of excess energy to oxygen, which could otherwise lead to the formation of harmful ROS.

Based on our calculations, we demonstrate that 2DCD spectroscopy offers more detailed spectral information compared to 2DES. However, it is important to note that the 2DCD signal is inherently weaker than that observed in 2DES, posing challenges for experimental implementation. Enhancement of the spectroscopic signal strengths is necessary for capturing the transient chiral dynamics described in this work. One of the effective strategies for signal enhancement can be manipulation of the pulse profile and phase during the photoexcitation process, which influences the light-matter interaction. We propose utilizing vector or vortex beams on ultrafast timescales to modulate and amplify the chiral contributions in 2DCD spectroscopy, a technique that has gained recent attention\cite{Mukamel1, Mukamel2, Mukamel3, Mukamel4}. We envision that 2DCD spectroscopy could be a potent tool for detecting structural changes associated with magnetic transition dipoles, suggesting its potential application in probing molecular structural alterations within protein complexes. Examples include studying photo-isomerization processes in protein complexes\cite{Miller Nature Chem 2015} and analyzing bond-breaking and dissociation dynamics of specific molecules\cite{Jieyang Science 2018, MingFu Lin Nature Chem 2018}. 

It is also of considerable interest to investigate charge transfer dynamics in PSII RC and BRC complexes. One of the main challenges in studying charge transfer dynamics is the weak nature of electronic transitions associated with these processes. Charge-transfer states often involve small changes in the electronic distribution over a relatively large distance within the molecular system. As a result, these states typically have low oscillator strengths, making them difficult to detect using conventional spectroscopic techniques, which rely on strong electronic transitions to provide detectable signals. However, the combination of electronic and magnetic transition dipoles inherent in 2DCD spectroscopy provides a promising platform for examining charge-transfer dynamics, an area that remains largely unexplored.

\section*{Conclusions} 

In conclusion, while the BRC and the PSII RC share structural similarities, they exhibit notable differences in chiral dynamics as observed through 2DCD spectroscopy. The BRC shows more distinct separation of main and cross peaks in both 2DES and 2DCD spectra, a feature attributed to its specific pigment arrangement and stronger excitonic couplings. In contrast, the PSII RC has less pronounced excitonic interactions. Analysis of chiral dynamics reveals that the BRC maintains stronger chiral signatures and longer coherence times, often exceeding 1 ps, highlighting more robust interactions between its pigments. Conversely, the PSII RC shows a faster decay of coherence, likely due to the symmetric yet less complex arrangement of key pigments such as $Chl_{D1}$ and $Chl_{D2}$. Notably, the 2DCD spectra, particularly the cross peaks, offer insights into the spatial organization and structural arrangement of cofactors within protein complexes, details that are not observable in 2DES. This comparative analysis of BRC and PSII RC provides valuable insights that might be linked to the evolutionary adaptations of photosynthetic complexes to thrive in diverse environments. Understanding these evolutionary strategies not only sheds light on the history of photosynthesis but also provides a foundation for bioengineering and designing artificial photosynthetic systems that could mimic these natural processes for sustainable energy solutions. More generally, 2DCD spectroscopy, which includes magnetic transition dipoles, offers a unique and powerful method for tracking structure-related dynamics in both natural and synthetic photosynthetic systems.



\section*{Methods}
\subsection{System-Bath Hamiltonian}

In the open quantum system, the total Hamiltonian can be written as four parts  
\begin{equation} 
\label{eq:Htot}
H = H_S + H_B + H_{SB} + H_{Ren},  
\end{equation}
where $H_{S}$, $H_{B}$ and $H_{SB}$ are the system, bath and system-bath interaction terms, respectively. The $H_{Ren}$ is the reorganization energy. The system and bath Hamiltonian can be written as 
\begin{equation}
\label{eq:HS}
H_S = \sum \limits_{m=1}^{N} \epsilon_{m} \ket{m} \bra{m} + \sum \limits_{m=1}^{N} \sum \limits_{n<m} J_{n,m} (\ket{m} \bra{n} + \ket{n} \bra{m} ),
\end{equation}
\begin{equation}
\label{eq:HB}
H_B = \sum \limits_{m=1}^{N} \sum \limits_{j=1}^{N_{b}^{m}} \left(\frac{p_{mj}^{2}}{2} + \frac{1}{2}\omega_{mj}x_{mj}^{2}\right) , 
\end{equation}
Where $\epsilon_{m}$ and $J_{n,m}$ is the m-th site energy and the excitonic coupling between n- and m-th pigments, respectively. $N$ is the total number of pigments. $N^{m}_{b}$ is the total number of bath modes coupled to the m-th molecule. $x_{mj}$ and $p_{mj}$ are the mass weighted position and momentum of j-th harmonic oscillator bath mode with frequency $\omega_{mj}$ in $H_{B}$. $H_{SB} = \sum_{m} K_m \Phi_m (x) $ is the interaction term and it describes the coupling between system and bath. It is assumed that $K_m$ only acts on system subspace and $\Phi_m (x)$ only on the bath degrees of freedom. In the following, we further assume a linear relation between bath coordinates and the system. The system-bath interaction is then given by  
\begin{equation}
\label{eq:HSB}
H_{SB} = \sum \limits_{m} K_m \sum \limits_{j} c_{mj} x_{mj}. 
\end{equation} 
We further restrict our considerations to pure electronic dephasing only, {\em i.e.} $K_m = a_m^{\dagger}a_m$. The reorganization energy is given as 
\begin{equation}
\label{eq:HREN}
H_{Ren} = \sum \limits_{m} \sum \limits_{j} K_m^{2} \frac{c_{mj}^{2}}{m_{mj}\omega_{mj}^{2}}. 
\end{equation}
This term compensates for artificial shifts of the system frequencies due to the system-bath interaction. The influence of the bath can be fully described by the spectral density, which show the form 
\begin{equation}
\label{eq:spectral density}
J_m(\omega) = \pi \sum \limits_{j} \frac{c_{mj}}{2m_{mj}\omega_{mj}} \delta(\omega - \omega_{mj})  = 2 \lambda \frac{\omega \gamma}{\omega^2 + \gamma^2}, 
\end{equation} 
where $\lambda$ is the reorganization energy and $\gamma$ is the high-frequency cutoff.  In this paper, we take the Lorentzian type of spectral density with Drude cutoff frequency. We assume the bath at each monomer to be independent but with identical spectra, {\em i,e.,} $J_n (\omega) = J_m (\omega)$.  

\subsection{Redfield equation of motion} 
The Redfield equation describes the time evolution of the density matrix of a system coupled to its environment. Under the Markov approximation, they provide a set of coupled master equations for the reduced density matrix of the system. In this form, we can accurately describe the dynamics of a dissipative system if the noise-inducing environment is weakly coupled and has short correlation times.The exact dynamics of the reduced density matrix $\sigma(t)$ of the system is given by the Nakajima-Zwanzig equation (simplified below with $\hbar$ = 1). 
\begin{equation}
\label{eq:Nakajima-Zwanzig}
\frac{\partial \sigma(t)}{\partial t} = -i \mathcal{L} \sigma(t) - \int_{0}^{t} d \tau  \mathcal{G}(\tau) \sigma(t-\tau), 
\end{equation}	
and  
\begin{equation}
\label{eq:G}
\mathcal{G}(\tau) = \text{tr}_B\{\mathcal{L}_{SB} e^{-i\mathcal{L}(1-P)\tau} \mathcal{L}_{SB} \rho_B  \}. 
\end{equation}
The Nakajima-Zwanzig equation is formally exact. However, since ${G}(\tau)$ is difficult to obtain, the Nakajima-Zwanzig equation has little practical use and can only serve as the basis for a more approximate theory. In order to obtain an equation that is easier to understand numerically, Redfield employs two approximations, the ``Born approximation''\cite{Born} and ``Markov approximation''\cite{Markov}, which gives the form 
\begin{equation}
\label{eq:Born approximation}
\mathcal{G}(\tau) \simeq \text{tr}_B\{\mathcal{L}_{SB} e^{-i (\mathcal{L}_S + \mathcal{L}_B) \tau} \mathcal{L}_{SB} \rho_B  \}. 
\end{equation} 
\begin{equation}
\label{eq:Markov approximation}
\sigma(t-\tau) \simeq e ^{i \mathcal{L}_S \tau} \sigma (t),  
\end{equation} 
Using Eqs.\ \ref{eq:Born approximation} and \ref{eq:Markov approximation}, we obtain the local-time motion equation of the reduced density matrix, which show the form 
\begin{equation}
\label{eq:Equations of motion}
\frac{\partial \sigma(t)}{\partial t} = -i [H_S, \sigma(t)] - \int_{0}^{t} d \tau  \text{tr}_B \{ [H_{SB}, [H_{SB} (\tau) , \sigma (t)\rho_B]]\},   
\end{equation}
Then, by introducing the eigenstates of the system $H_S \ket{a} = E_a \ket{a}$. The original form of the Redfield equation can be obtained 
\begin{equation}
\label{eq:Redfield}
\quad\quad
\frac{\partial \sigma_{\mu \nu}(t)}{\sigma t} = -i \omega_{\mu \nu} \sigma_{\mu \nu} (t) + \sum_{\kappa \lambda} R_{\mu \nu \kappa \lambda}\sigma_{\kappa \lambda}(t).
\end{equation} 
$\sigma_{\mu \nu}(t)= \bra{\mu} \sigma(t) \ket{\nu}$ denotes the matrix elements of the reduced density matrix in the eigenstate representation, and $\omega_{\mu \nu} = E_{\mu} - E_{\nu}$ are the eigen-frequencies of the system.

\subsection{Light-matter Interaction term}
The light-matter interaction Hamiltonian is given by 
\begin{equation}
\label{eq:H_int}
H_{int} =  -\boldsymbol{\mu} \cdot \boldsymbol{E}(t) -\boldsymbol{m} \cdot \boldsymbol{B}(t) -\boldsymbol{q} \cdot \nabla \boldsymbol{E}(t) \,, 
\end{equation}
in this context, $\boldsymbol{\mu}$ denotes the transition electric dipole moment, $\boldsymbol{m}$ denotes the transition magnetic dipole moment, and $\boldsymbol{q}$ denotes the quadrupole moment.  $\boldsymbol{E}(t)$ and $\boldsymbol{B}(t)$ represent the electric and magnetic fields, respectively, while $\nabla$ indicates the gradient operator. 

\subsection{Linear spectroscopy} 
We utilize the first-order response function to compute the absorption and circular dichroism spectra. Based on the field-matter interaction from Eq.\ \ref{eq:H_int}, the linear signal can be expressed as follows 
\begin{equation}
\begin{split}
\label{eq:s1 chir}
S^{(1)}(\Gamma) = \int dt dt_1  (
&R^{(1)}_{\mu\mu}(t_1) \cdot (\boldsymbol{E}_S(t) \otimes \boldsymbol{E}_1(t-t_1)) + 
R^{(1)}_{m\mu}(t_1) \cdot  (\boldsymbol{B}_S(t) \otimes \boldsymbol{E}_1(t-t_1))  + \\ 
&R^{(1)}_{\mu m}(t_1) \cdot (\boldsymbol{E}_S(t) \otimes \boldsymbol{B}_1(t-t_1))\,, 
\end{split}
\end{equation} 
and
\begin{equation}
\begin{split}
\label{eq:R mumu}
R^{(1)}_{\mu\mu}(t_1) = i \langle \boldsymbol{\mu} G(t_1) \boldsymbol{\mu}^{\times}\rho(-\infty) \rangle\,,  
\end{split}
\end{equation}
\begin{equation}
\begin{split}
\label{eq:R mmu}
R^{(1)}_{m\mu}(t_1) = i \langle \boldsymbol{m} G(t_1) \boldsymbol{\mu}^{\times}\rho(-\infty)  \rangle\,,  
\end{split}
\end{equation} 
respectively. Here, $\Gamma$ represents the set of parameters that govern the signal (including field-matter interaction, laser pulse polarization, wave vector configuration, central frequencies, and bandwidths). The symbols $\boldsymbol{E}_1$ and $\boldsymbol{B}_1$ denote the electric and magnetic fields of the incoming beams, while $\boldsymbol{E}_S$ and $\boldsymbol{B}_S$ denote the electric and magnetic fields of the probe beams. $G(t)$ indicates the time-evolution operator. For the electric dipole moment operator, we show $\mu = \mu_+ + \mu_-$ with $\mu_+ = \sum \limits_{n=1}^{N} \mu_n \ket{n}\bra{0}$ and $\mu_- = \sum \limits_{n=1}^{N} \mu_n \ket{0}\bra{n}$. For the magnetic dipole moment, it shows $m = m_+ + m_-$, where $m_+ = \sum \limits_{n=1}^{N} iM_n \ket{n}\bra{0}$ and $m_- = \sum \limits_{n=1}^{N} -iM_n \ket{0}\bra{n}$. The $\mu_n$ and $M_n$ are real values. 

The linear absorption spectrum $I(\Omega)$ is obtained by Fourier transform of $S^{(1)} (\Gamma)$ over $t_1$ with the linearly polarized light. The circular dichroism (CD) spectrum, on the other hand, is expressed as the difference in signal between left and right circularly polarized light, and can be represented in the form
\begin{equation}
\begin{split}
\label{eq:las and CD}
I(\omega)= &\int_{-\infty}^{\infty} dt_1 e^{i\omega t_1} S^{(1)} (\uparrow,\Gamma)\,, \\ 
CD(\omega)=&\int_{-\infty}^{\infty} dt_1 e^{i\omega t_1} S^{(1)} (\circlearrowleft,\Gamma) 
- \int_{-\infty}^{\infty} dt_1 e^{i\omega t_1}  S^{(1)} (\circlearrowright,\Gamma)\,, 
\end{split}
\end{equation}
where $\uparrow$, $\circlearrowleft$ and $\circlearrowright$ denote the linearly, left and right circularly polarized light, separately.

\subsection{Nonlinear spectroscopy} 
We employ the third-order response function to calculate the nonlinear spectroscopy. The expression of nonlinear signal can be written as 
\begin{equation}
\begin{split}
\label{eq:S3}
S^{(3)} (\Gamma) = &\int dt dt_3 dt_2 dt_1 (R^{(3)}_{\mu\mu\mu\mu} (t_3,t_2,t_1)  \cdot (\boldsymbol{E}_S (t) \otimes \boldsymbol{E}_3  \otimes \boldsymbol{E}_2  \otimes\boldsymbol{E}_1 )+ \\
&R^{(3)}_{m\mu \mu  \mu} \cdot (\boldsymbol{B}_S \otimes \boldsymbol{E}_3 \otimes \boldsymbol{E}_2 \otimes \boldsymbol{E}_1 ) + R^{(3)}_{\mu m \mu  \mu} \cdot (\boldsymbol{E}_S \otimes \boldsymbol{B}_3 \otimes \boldsymbol{E}_2 \otimes \boldsymbol{E}_1 ) + \\
&R^{(3)}_{\mu \mu m \mu} \cdot (\boldsymbol{E}_S \otimes \boldsymbol{E}_3 \otimes \boldsymbol{B}_2 \otimes \boldsymbol{E}_1 ) + R^{(3)}_{\mu \mu \mu m} \cdot (\boldsymbol{E}_S \otimes \boldsymbol{E}_3 \otimes \boldsymbol{E}_2 \otimes \boldsymbol{B}_1 ) ) \,,
\end{split} 
\end{equation}
with
\begin{equation}
\begin{split}
\label{eq:R3 mumumumu}
R^{(3)}_{\mu\mu\mu\mu}(t_3,t_2,t_1) = -i \langle \boldsymbol{\mu} G(t_3)
\boldsymbol{\mu}^{\times} G(t_2)\boldsymbol{\mu}^{\times} G(t_1)\boldsymbol{\mu}^{\times} \rho(-\infty) \rangle \,, 
\end{split}
\end{equation}
\begin{equation}
\begin{split}
\label{eq:sysham}
R^{(3)}_{\mu\mu\mu m}(t_3,t_2,t_1) &= -i \langle \boldsymbol{\mu} G(t_3)
\boldsymbol{\mu}^{\times} G(t_2)\boldsymbol{\mu}^{\times} G(t_1)\boldsymbol{m}^{\times} \rho(-\infty) \rangle \,.
\end{split} 
\end{equation}
$t_1$ is coherence time, $t_2$ is waiting time, and $t_3$ is detection time. 

The 2DES is calculated by Fourier transform of  $S^{(3)} (\Gamma)$ in Eq.\ \ref{eq:S3} over the coherence time and the rephasing time. 
\begin{equation}
\begin{split}
\label{eq:S 2des}
S_{2DES}(\omega_t,T,\omega_{\tau})= &\int_{-\infty}^{\infty} d\tau e^{i\omega_{\tau} \tau} \int_{-\infty}^{\infty} dt e^{i\omega_t t} S^{(3)} (\uparrow,\Gamma),
\end{split}
\end{equation}
the 2DCD spectra is defined as the difference between two nonlinear signals from left- and right-circularly polarized pulses, which shows 
\begin{equation}
\begin{split}
\label{eq:S 2dcd}
S_{2DCD}(\omega_t,T,\omega_{\tau})= &\int_{-\infty}^{\infty} d\tau e^{i\omega_{\tau} \tau} \int_{-\infty}^{\infty} dt e^{i\omega_t t} S^{(3)} (\circlearrowleft,\Gamma)-\int_{-\infty}^{\infty} d\tau e^{i\omega_{\tau} \tau} \int_{-\infty}^{\infty} dt e^{i\omega_t t} S^{(3)} (\circlearrowright,\Gamma)\,.
\end{split}
\end{equation}
The detailed model and method information is provided in the Supporting Information (SI).

%
\begin{addendum}
\item This work was supported by NSFC Grant with No.\ 12274247, Yongjiang talents program with No.\ 2022A-094-G, Ningbo International Science and Technology Cooperation with No.\ 2023H009, ‘Lixue+’ Innovation Leading Project and the foundation of national excellent young scientist. The Next Generation Chemistry theme at the Rosalind Franklin Institute is supported by the EPSRC (V011359/1 (P)) (AJ). 

\item[Supporting information] The detailed information of Model Hamiltonian, light-matter interaction, linear spectroscopy, the traces of selected peaks and population-transfer-related chiral dynamics evolving with time. 

\item[Competing Interests] The authors declare that they have no competing financial interests. 

\item[Correspondence] Correspondence of paper should be addressed to A. J. (email: Ajay.Jha@rfi.ac.uk) and H. -G. D. ~(email: duanhongguang@nbu.edu.cn). 

\item[Author Contributions] H. -G. D. and A. J. conceived the research project. Y.L.Y., Z. H. L. calculated the 2DES and 2DCD spectra. Y.L.Y., Z. H. L. analyzed the data with the help from F. Z., P.-P. Z., H.-X. H., A. J. and H. -G. D.. H. -G. D. and A. J. wrote the first draft the manuscript and all authors contributed towards the refinement of the manuscript. H. -G. D. supervised this project. 

\end{addendum}
%
\newpage

\newpage

\newpage

\newpage

\newpage


\end{document}